# Shift of the $2_1^+$ state of $^{10}$Be in ternary cold fission of $^{252}$Cf

Ş. Mişicu,[1,2,*] A. Săndulescu,[1,2] and W. Greiner[2]

[1]*National Institute for Nuclear Physics, Bucharest-Magurele, P.O. Box MG-6, Romania*
[2]*Institut für Theoretische Physik der J. W. Goethe Universität, Frankfurt am Main, Germany*
(Received 2 December 1999; published 17 March 2000)

Recent experimental data indicate that in the ternary cold fission of $^{252}$Cf the energy of the first excited state of the accompanying light cluster $^{10}$Be is decreased by an amount ranging between $\approx 6$ and 26 keV. A model is proposed to calculate the shift of the vibrational $2_1^+$ state in $^{10}$Be when its heavy companions are the even-even nuclei $^{146}$Ba and $^{96}$Sr. The stiffness parameters of the $\beta$ vibrations are calculated within the self-consistent Hartree-Fock method with BCS pairing correlations taken into account, and their change is determined by the interaction of the light cluster with the heavy fragments. The results point to a dependence of the shift magnitude and signature on the relative distance between the three clusters and their mutual orientation. Eventually it is the anharmonic perturbation of the spherical vibrator which is responsible for obtaining a negative energy shift of the $2_1^+$ state.

Recently, the spontaneous ternary fission of $^{252}$Cf, accompanied by $^{10}$Be in a rather excited state, has been observed in two different experiments. In the first one, the prompt $\gamma$-ray emission was investigated in the spontaneous hot fission of $^{252}$Cf, performed with the aid of the $4\pi$ Darmstadt-Heidelberg Crystal Ball spectrometer [1]. An interesting observation was that some of the $\gamma$ quanta associated with the deexcitation of the first excited level in $^{10}$Be apparently are emitted from a resting source. It was concluded that the lifetime of the $2_1^+$ level in $^{10}$Be, which is 125 fs according to [2], should be smaller than the lifetime of the ternary fission mode. It was also noted that the existence of a ternary scission barrier, which is penetrated by the three fragments after such a large time, is linked with the existence of a molecular-type of nuclear structure.

Later on a similar observation has been made for the cold limit of the same spontaneous decay process with the Gammasphere consisting of 72 detectors [3]. The $\gamma$ ray from the first $2^+$ state in $^{10}$Be, observed in coincidence with the $\gamma$ rays of the fission partners $^{146}$Ba and $^{96}$Sr, exhibits no Doppler shift. In addition, it was observed that the energy value of the $2_1^+$ state in such a quasibound configuration is lowered by 6 keV compared to the value compiled for the free configuration ($E_\gamma = 3368.03$ keV) [4]. Very recently the value compiled by Ajzenberg-Selove was confirmed experimentally, with a slight modification ($E_\gamma = 3368.34 \pm 0.43$ keV), using the reaction $^9$Be$(d,p\gamma)^{10}$Be at $E_d = 1.0$ MeV [5]. After reanalyzing the data, obtained with the Gammasphere, more splittings were discovered, e.g., $^{99}$Y+$^{142}$Cs and $^{108}$Mo+$^{134}$Te, with energy shifts 16 keV and 26 keV, respectively [6]. The enhancement of the shift when moving towards sphericity was qualitatively motivated by the largest overlap of the $^{10}$Be nucleus with the spherical fragment, which has the tendency to decrease the average spacing $\hbar\omega$ of the harmonic oscillator shells.

The scope of this Rapid Communication is to attempt an explanation of the above recent experimental investigations on the shift of the $2_1^+$ state in $^{10}$Be using a molecular model for the ternary cold fission as presented earlier [7,8]. According to the molecular scenario, the ternary cold fission of $^{252}$Cf is a process consisting in the decay of a quasibound molecular structure in which the heavier fragments are almost collinear and the light particle (e.g., $\alpha$, $^3$Li, $^{10}$Be, or $^{12}$C), which is responsible for the quasimolecular bonding, is orbiting in the neighborhood of the equatorial region. This is similar to the case encountered in molecular physics, where, in a linear or nonlinear chain of three atoms, the central atom ensures two bondings with the eccentric atoms [9]. Although we deal with a dynamical process, we adopt in this paper an adiabatic approach, such that the energy shift will not depend on the history of the decay process and will adjust slowly to the value of the main fission variable, i.e., the elongation of the dinuclear subsystem, formed from two heavier clusters. For a given configuration of the three clusters composing the giant nuclear molecule, we write the static energy as the sum of the deformation energies and their mutual potentials:

$$E_{\text{ternary}} = \sum_{i=1}^{3} E_{\text{def}}(\{\beta\}_i) + \frac{1}{2}\sum_{i\neq j}^{3} V_{ij}(r_{ij},\{\beta\}_i,\{\beta\}_j). \quad (1)$$

By $r_{ij}$ we denote the reciprocal distance between the clusters $i$ and $j$, and $\{\beta_i\}$ represents the deformation of cluster $i$, whose multipolar components are the quadrupole $\beta_2$, octupole $\beta_3$, and hexadecupole $\beta_4$ deformations. The above ansatz is similar to the total energy of the system at the scission point considered in Ref. [10] for binary fission. The reference is made with respect to the energy of three infinitely distant fragments in their ground states.

In order to compute the deformation energy, we resort to the self-consistent Hartree-Fock method with BCS pairing correlations (HF+BCS) following the scheme presented in Refs. [11,12] for a free nucleus. For that we first determine the equilibrium deformation by seeking the HF minimum and then compute the deformation energy curve by constraining its quadrupole moment. In principle one should also constrain over higher multipole moments ($Q_3, Q_4, \ldots$). In this paper we make the assumption that for a given value of the quadrupole moment $Q_2$, the octupole and hexadecupole moments $Q_3$, $Q_4$, respectively, have the same values as those obtained upon minimization. In what follows we employ as deformation variable for the $i$th cluster the quadrupole deformation $\beta_{2i} \equiv \beta_i (i=1,2,3)$. The deformation energy for the light cluster 3 resembles very much a vibrator with a small cubic anharmonicity for not too large departures from the spherical equilibrium position:

$$E_{\text{def}}(\beta_3) = \frac{1}{2} C_2 \beta_3^2 + C_3 \beta_3^3, \tag{2}$$

where the stiffness parameters have the numerical values, $C_2 = 7.688$ MeV and $C_3 = -2.855$ MeV, obtained upon interpolating the Hartree-Fock deformation curve.

*Electronic address: misicu@theor1.theory.nipne.ro

Then, for a given configuration, which should correspond to a precise point on the fission path, the static energy of the quasimolecule must have a minimum with respect to variations in the deformations [10], i.e.,

$$\frac{\partial E_{\text{ternary}}}{\partial \beta_i} = 0, \qquad (i=1,2,3). \tag{3}$$

Since we are interested only in determining the collective spectrum of the $^{10}$Be subsystem of the quasibound system $^{252}$Cf, we write its Hamiltonian in the form

$$H_3^{(\text{qb})} = H_3^{(\text{free})} + V_{13}(\mathbf{r}_{13}, \beta_1, \beta_3) + V_{23}(\mathbf{r}_{23}, \beta_2, \beta_3), \tag{4}$$

where the exponent (qb) stands for the quasibound case, whereas (free) denotes the case when $^{10}$Be is not under the influence of external forces.

In the free case, the quantized form of the third cluster Hamiltonian reads

$$\hat{H}_3^{(\text{free})} = -\frac{\hbar^2}{2B} \frac{\partial^2}{\partial \beta_3^2} + E_{\text{def}}(\beta_3), \tag{5}$$

where the effective mass parameter is computed from the experimental values of $B(E2, 0_1^+ \to 2_1^+)$ [13] and the energy of the $2_1^+$ state [14]

$$B = \frac{5 \left( \frac{3Z_3 e R_0^2}{4\pi} \right)^2 \hbar^2}{2 \hbar \omega_{2_1^+} B(E2, 0_1^+ \to 2_1^+)}. \tag{6}$$

When the interaction between the light cluster, 3, and the heavy clusters, 1 and 2, is turned on, the effective deformation energy (2) will be modified. To make the problem tractable we expand this interaction with respect to the deformation of the third cluster $\beta_3$, keeping at the same time the deformations of the heavier cluster, $\beta_1$ and $\beta_2$, freezed. For small values of the deformation, using the mimization condition (3), we get

$$V_{13}(\beta_3) + V_{23}(\beta_3) \approx \left( 1 + \frac{1}{2!} \beta_3^2 \frac{\partial^2}{\partial \beta_3^2} + \frac{1}{3!} \beta_3^3 \frac{\partial^3}{\partial \beta_3^3} \right) \times [V_{13}(0) + V_{23}(0)]. \tag{7}$$

In the above equation, we omitted the dependence on the intercluster distance and the deformation of the heavier partner, which enter as parameters. Accordingly, the terms multiplying the square and the cube of $\beta_3$ in the above expansion have to be added to the stiffness parameters $C_2$ and $C_3$, respectively, entering in Eq. (2), in order to get the expression of the anharmonic vibrational Hamiltonian in the quasibound configuration. We are then left to derive the spectrum of a one-dimensional harmonic oscillator with a small cubic anharmonic perturbation. Applying the stationary perturbation theory in the second-order approximation [15], we obtain, for the difference between the energy of the $2_1^+$ state and the ground state, the expression

$$\hbar \omega_{2_1^+}^{(\text{qb})} = \hbar \omega \left\{ 1 - \frac{15}{2} \frac{C_3'^2}{\hbar \omega} \left( \frac{\hbar}{B \omega} \right)^3 \right\}, \tag{8}$$

where $\hbar \omega$ corresponds to the energy level difference for the unperturbed harmonic oscillator, i.e.,

$$\hbar \omega = \sqrt{\frac{C_2'}{B}}, \tag{9}$$

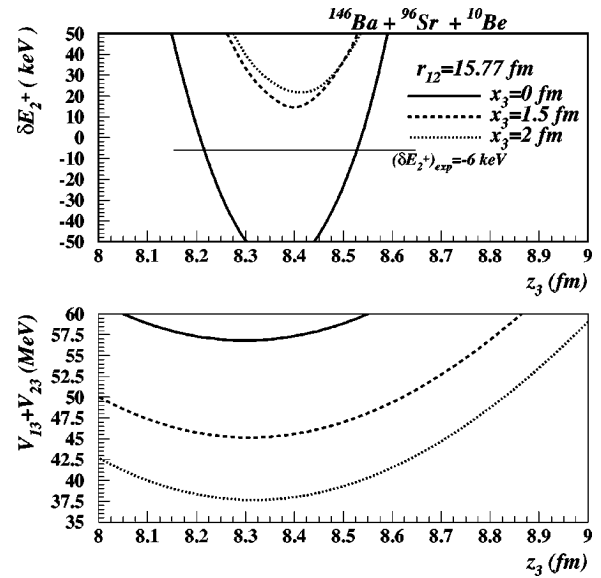

FIG. 1. The energy shift of the $2_1^+$ state (upper panel) and total potential energy ($V_{13} + V_{23}$) of $^{10}$Be (lower panel) as a function of the $z$-component of the position of the third cluster. The origin of the coordinate system coincides with the center of mass of cluster 1. The above mentioned dependence is plotted for three values of $x_3$: 0 fm (full line), 1.5 fm (dashed line), and 2 fm (dotted line) and the fixed distance between clusters 1 and 2: $r_{12} = 15.77$ fm. It is worthwhile to notice the location of the minima of the two plotted quantities at almost the same value of $z_3$. The experimental value of the energy shift is sketched by a thin straight line at $-6$ keV.

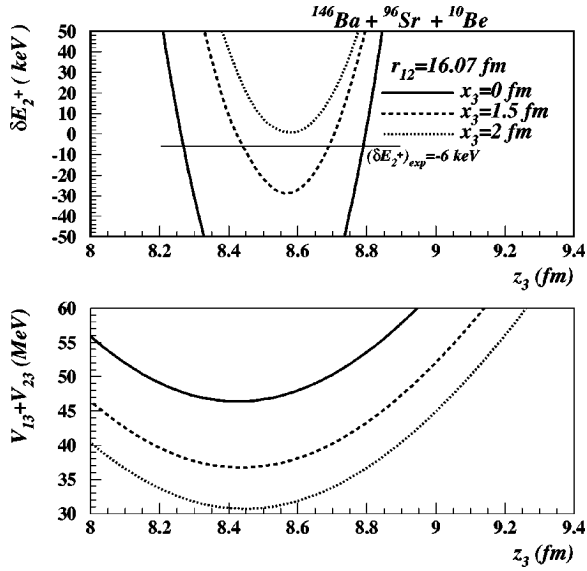

FIG. 2. Same as in Fig. 1 for $r_{12}=16.07$ fm. In this case, the energy shift has the most pronounced tendency to acquire negative values.

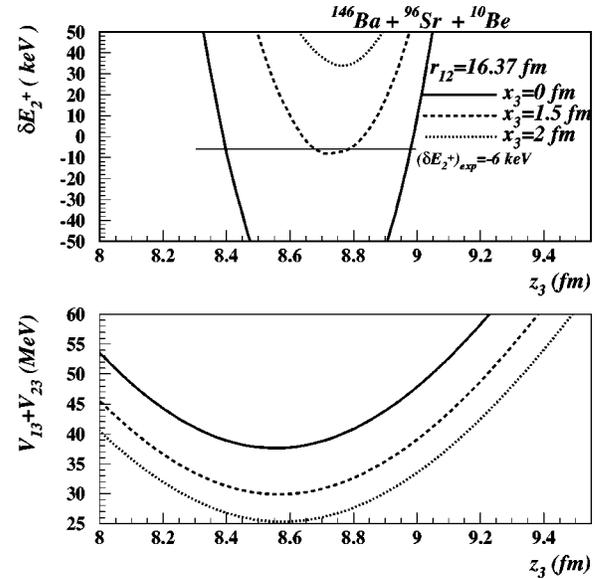

FIG. 3. Same as in Fig. 1 for $r_{12}=16.37$ fm.

$C'_{2,3}$ being the modified stiffness parameters of the quadratic and cubic potential terms. Then, the energy shift $\delta E_{2^+}$ of the first $2^+$ state is obtained by simply subtracting from $\hbar\omega_{2_1^+}^{(qb)}$ the quantum energy of the free, noninteracting case, which has the same expression as Eq. (8), $C'_{2,3}$ being traded for $C_{2,3}$.

In Figs. 1–3 we represented, for different distances between the heavy clusters, $r_{12}$, the energy shift of the $2_1^+$ state, in the case of the splitting $^{146}$Ba+$^{96}$Sr+$^{10}$Be, as a function of the location of the light cluster on the molecular $z$ axis (upper panel). In the lower panel, the sum $V_{13}+V_{23}$ of the interactions between clusters 1 and 3 and between clusters 2 and 3, is plotted. The heavy fragments were supposed to have their symmetry axes aligned. An interesting fact, which becomes apparent from the inspection of these figures, is the close proximity between the location of the minima of the two sets of curves, which may suggest that the lowest value of the energy shift is obtained in the case where the light cluster is located on, or very close to, the electronuclear saddle curve. In Ref. [16], the electronuclear saddle point was assigned to the position on the fission (molecular) axis where the combined Coulomb and nuclear forces exerted by the heavy fragments on the light cluster cancel each other, and the potential along the same axis has relative minima. Moving off the fission axis, we defined the electronuclear saddle curve as the geometrical locus of points where the $z$ component of the force exerted by cluster 1 on cluster 3 is compensated by the forces exerted by cluster 2 on cluster 3 [7]. The figures represent the calculations for tip distances between the heavy clusters, $d=1.70$, 2, and 2.30 fm, which translated into intercluster distances are $r_{12}=15.77$, 16.07 and 16.37 fm. Outside this range the negative energy shifts are gradually disappearing. If one makes the assumption that the light cluster is located on the bottom of the potential pocket, or, in quantum terms, is filling up the first state of the quantum well produced by the interaction of the three fragments, then the experimental shift, $\delta E_{2^+}=-6$ keV, can be reproduced if the tip distance between the heavy fragments is confined in the interval 1.5–2.5 fm, and the cluster $^{10}$Be is located off the fission axis at a height $x_3$ not larger than 2 fm. Such a configuration is compatible with the light-particle trajectory in the so-called adiabatic scenario of the ternary cold fission obtained previously by us [7]. In this scenario the light cluster is in touch with each heavy fragment, the touching point being defined such that the densities of two nuclei in contact are half of their central value. Expressed in more pedestrian terms, we expect that a polarization of $^{10}$Be, consistent with the observed 6 keV shift, arises in a triangular configuration of the giant molecule rather than a linear one. To specify even better this position one should undertake a tunneling path calculation of the type proposed in Ref. [7] for the $\alpha$ cluster. It would then be natural to make the suggestion that the shift of the $\gamma$ quanta, whose lifetime can be easily deduced upon comparison with the free (not shifted) case, is a probe for the decay time of the nuclear molecule.

In conclusion, our study, based on a molecular scenario in which the smaller cluster is orbiting in the equatorial region of the dinuclear subsystem formed by the heavier fragments, points to a negative energy shift of the first $2^+$ state in $^{10}$Be, only for a narrow range of the distance between the two heavy clusters, which at their turn are in a slight overlap or in touch with the third, light cluster. Essential in our final result was the account of the cubic anharmonicities in the vibrational Hamiltonian. Repeating the calculations, and keeping only the terms quadratic in deformation, no negative shift was obtained.

[1] P. Singer, Yu. Kopach, M. Mutterer, M. Klemens, A. Hotzel, D. Schwalm, P. Thirolf, and M. Hesse, in *Proceedings of the 3rd International Conference on Dynamical Aspects of Nuclear Fission*, Časta Papiernička, Slovakia, 1996, edited by J. Kliman and B. Pustylnik (Dubna, 1996) p. 262.

[2] Electronic retrieval from http://www.nndc.bnl.gov/nndcscr/testwww/AR010BE.HTML

[3] A.V. Ramayya, J.K. Hwang, J.H. Hamilton, A. Săndulescu, A. Florescu, G.M. Ter-Akopian, A.V. Daniel, G.S. Popeko, W. Greiner, J.O. Rasmussen, M.A. Stoyer, J.D. Cole, and GANDS Collaboration, Phys. Rev. Lett. **81**, 947 (1998).

[4] F. Ajzenberg-Selove, Nucl. Phys. **A490**, 1 (1988).

[5] B. Burggraf, K. Fazrin, J. Grabis, Th. Last, E. Manthey, H.P. Trautvetter, and C. Rolfs, J. Phys. G **25**, L71 (1999).

[6] J.H. Hamilton, A.V. Ramayya, J.K. Hwang, W. Greiner, J. Kormicki, G.M. Ter-Akopian, A.V. Daniel, Yu.Ts. Oganessian, J.D. Cole, and GANDS95 Collaboration (unpublished).

[7] A. Săndulescu, Ş. Mişicu, F. Carstoiu, and W. Greiner, Phys. Part. Nuclei **30**, 386 (1999).

[8] S. Mişicu, P.O. Hess, A. Săndulescu, and W. Greiner, J. Phys. G **25**, L147 (1999).

[9] E.B. Wilson, J.C. Decius, and P.C. Cross, *Molecular Vibrations* (McGraw-Hill, New York, 1955).

[10] F. Dickman and K. Dietrich, Nucl. Phys. **A129**, 241 (1969).

[11] D. Vautherin, Phys. Rev. C **7**, 296 (1973).

[12] P. Quentin and H. Flocard, Annu. Rev. Nucl. Part. Sci. **28**, 523 (1978).

[13] S. Raman, C.H. Malarkey, W.T. Milner, C.W. Nestor, Jr., and P.H. Stelson, At. Data Nucl. Data Tables **36**, 1 (1987).

[14] C.Y. Wong, Nucl. Data, Sect. A **4**, 271 (1968).

[15] S. Flügge, *Practical Quantum Mechanics I* (Springer-Verlag, Berlin, 1971).

[16] Ş. Mişicu, A. Săndulescu, F. Carstoiu, M. Rizea, and W. Greiner, Nuovo Cimento A **112**, 313 (1999).